\documentclass[10pt,prl,aps,twocolumn,showpacs,superscriptaddress]{revtex4}
\usepackage{graphicx}
\usepackage{amssymb}
\usepackage{bm}

\newcommand{\bc}{\begin{center}}
\newcommand{\ec}{\end{center}}
\def\ba#1{\begin{array}{#1}\displaystyle}
\newcommand{\ea}{\end{array}}

\newcommand{\beq}{\begin{equation}}
\newcommand{\eeq}{\end{equation}}
\newcommand{\beqa}{\begin{eqnarray}}
\newcommand{\eeqa}{\end{eqnarray}}

\newcommand{\n}{\nonumber\\}
\newcommand{\bi}{\begin{itemize}}
\newcommand{\ei}{\end{itemize}}

\def\lt#1{\left#1}
\def\rt#1{\right#1}

\def\h#1{\hat{#1}}
\def\b#1{\bar{#1}}
\def\frc#1#2{\frac{#1}{#2}}

\newcommand{\p}{\partial}

\newcommand{\vac}{{\rm vac}}
\newcommand{\bra}{\langle}
\newcommand{\ket}{\rangle}

\newcommand{\R}{{\mathbb{R}}}

\newcommand{\Tr}{{\rm Tr}}

\newcommand{\Or}{{\cal O}}

\newcommand{\ep}{\epsilon}

\begin{document}

\title{Bi-partite entanglement entropy in massive two-dimensional quantum field theory}

\author{Benjamin Doyon}
\affiliation{Department of mathematical sciences, Durham University, DH1 3LE, United Kingdom}

\date{\today}


\begin{abstract}
Recently, Cardy, Castro Alvaredo and the author obtained the first
exponential correction to saturation of the bi-partite
entanglement entropy at large region length, in massive two-dimensional integrable quantum field theory.
It only depends on the
particle content of the model, and not on the way particles scatter. Based on general analyticity arguments for form factors, we propose that this result
is universal, and holds for any massive two-dimensional model (also {\em out of integrability}). We suggest a link of this result with counting pair creations far in the past.
\end{abstract}

\maketitle

Entanglement is a fundamental characteristic of quantum mechanics: a measurement
at a point in space may affect instantaneously measurements performed elsewhere in a way that cannot be described
by local variables.
This leads for instance to enhanced performances of quantum algorithms.
Many measures of quantum entanglement between local degrees of freedom
in pure states are available, but
one that has been widely studied recently is the entanglement
entropy \cite{ent} (it appeared first in the context of black hole entropy \cite{bh}).
It measures the entanglement between two sectors of compatible observables, and is a powerful tool for assessing properties of ground states. Its
study has already led to a further understanding of quantum systems more generally.

We study the universal behaviour of the entanglement between two spacial regions, $A$ and its complement $\b{A}$, in the ground state of any extended one-dimensional systems. The Hilbert space is ${\cal H}=
{\cal H}_A \otimes {\cal H}_{\b{A}}$, and the entanglement entropy
is the von Neumann entropy of the reduced density
matrix associated to $A$: $S_A = - \Tr_{{\cal H}_A} \rho_A\log(\rho_A)$, $\rho_A = \Tr_{{\cal
H}_{\b{A}}} |{\rm gs}\ket\bra{\rm gs}|$ where $|{\rm gs}\ket$ is
the ground state. For a segment $A$ of length $R$, the universal scaling limit, near to a quantum critical point, is
obtained by making the
correlation lengths $\xi_\alpha,\, \alpha=1,\ldots,\ell$ and $R$
go to infinity with $R/\xi_\alpha=rm_\alpha$. This defines the dimensionful
length $r$ and a set of particle masses
$m_\alpha$ (with $m_1\leq m_\alpha \forall \alpha$) of a quantum field theory (QFT). We
will assume space-time translation and relativistic invariance, with discrete
mass spectrum (as occurs in most cases).

When the length $r$ becomes much larger than all $m_\alpha^{-1}$,
the entanglement entropy is known to saturate to a constant that
scales like $\frc{c}3 \log(\xi_1) + U$, where $c$ is the central
charge of the critical point and $U$ is a finite number. Beyond this saturation constant, one expects exponential corrections controlled by the masses. However, the exact form of these corrections is expected {\em a priori}, like the constant $U$ and the correlation lengths, to depend in a very non-trivial way on the details of the QFT.

In \cite{a}, then in a slightly more general situation in \cite{b},
the leading correction to saturation of the entanglement
entropy at large $r$ was evaluated in massive {\em integrable}
QFT, and found to take the same form independently of the
(factorised) scattering matrix: \beq\label{main}
    S_A = \frc{c}3 \log(\xi_1) + U - \frc18 \sum_{\alpha=1}^\ell K_0(2rm_\alpha) + O(e^{-3rm_1})
\eeq where $K_0$ is the modified Bessel function.

In this letter we propose that {\em the same leading correction holds generally in massive
QFT, not necessarily integrable}. This is a very general, universal
and surprising result, as no details of the scattering matrix are involved, and it only depends on the particle spectrum (a basic QFT property). It is also an {\em exact} low-energy result out of integrability. This is both notoriously difficult to obtain (it is impossible to calculate such exponential corrections for correlation functions out of integrabiliy), and closer to experimental situations than perturbative results. Indeed, because of asymptotic freedom, perturbation theory of renormalisable models only reliably provide high-energy information, whereas most experiments on solid state physics occur at energies much below the energy gap (this is at the basis of the success of massive integrable models, see e.g. \cite{ek}). Also, it has been observed in integrable models that leading large-distance corrections in general give good approximations up to very small distances, typically $rm_1 = 0.1$. Likewise, formula (\ref{main}) should give numerically a good approximation to the universal entanglement entropy in general QFT. At a deeper level, the leading correction contains ``clean'' information
about the particle spectrum, an important property of a QFT model. It gives all the masses
with $m_\alpha<3m_1/2$: ``half'' of the spectrum, since in non-integrable models $m_\alpha < 2m_1 \forall \alpha$
(otherwise the particle would decay).

We provide a derivation of this result based on the scattering theory of QFT, which we review, and a physical interpretation that explains some of its features.

{\bf Hilbert space and local fields.}

The Hilbert space of massive two-dimensional relativistic QFT is
formed by asymptotic states \footnote{An asymptotic state is a linear
combination of field configurations such that at times far in the
past or far in the future, it tends to a single configuration of freely propagating and
well separated extended wave packets.}. There are two
natural orthonormal bases: particles in the far past ($in$), and in the far future
($out$). They will be denoted by $|\theta_1,\ldots,\theta_k\ket_{\alpha_1,\ldots,\alpha_k}^{in,out}$
where $\theta_i,\,i=1,\ldots,k$
are rapidities and $\alpha_i,\,i=1,\ldots,k$ are quantum numbers
(with vacuum $|\vac\ket$). In order for
particles coming from the past to interact, they
must have decreasing rapidities from left to
right. Hence we will adopt the ordering
$\theta_1>\ldots>\theta_k$ for $in$ states, and the opposite
for $out$ states. The bases are fixed once impact parameters are chosen: extrapolated
trajectories meet, say, at $x=0,\,t=0$. The energy and momentum are
$\sum_i m_{\alpha_i} \cosh(\theta_i)$ and
$\sum_i m_{\alpha_i} \sinh(\theta_i)$, where $m_\alpha$ are the masses.

The Hilbert space alone is not enough to fix a QFT model.
We need to specify observables. The most important is the
energy field $\ep(x)$, with $H = \int_{-\infty}^\infty dx \ep(x)$
where $x$ is the position. Local
energy measurements are quantum mechanically independent at
space-like distances: $[\ep(x),\ep(x')]=0$ for $x\neq x'$. In QFT,
we look for all {\em local} fields $\Or(x)$, defined by $[\Or(x),\ep(x')]=0,\,x\neq x'$.

Another observable quantity is the scattering matrix (or
$S$-matrix): the amplitude of probability for a given $in$
configuration to end up as a given $out$ configuration.
The LSZ formula gives it in terms of correlation functions of local fields
$\Psi_{\alpha}(x)$ ``associated'' to the particles. But the $S$-matrix is believed to completely fix
a QFT model, hence the opposite also holds: local
fields can be deduced from it. Below we take this viewpoint.

{\bf The replica trick and branch-point twist fields}

We review the arguments of \cite{a,b}. There, the well-known replica trick $S_A = -\lim_{n\to1}
\frc{d}{dn} \Tr_{{\cal H}_A}(\rho_A^n)$ was used to evaluate the entanglement entropy. The quantity
$\Tr_{{\cal H}_A}(\rho_A^n)$ is identified in the scaling limit as
the QFT partition function on a Riemann
surface with $n$ sheets cyclicly connected on the interval $A$.

Let us consider instead a model composed of $n$ independent copies
of the initial QFT model on $\R^2$. The quantum
numbers become doublets $(\alpha,j)$, with $j=1,2,\ldots,n$
representing the copy number, and the local fields acquire an
index $j$. The total energy density is
$\ep(x) = \sum_j \ep_j(x)$. Particles on different copies do not
interact (trivial scattering):
$|\theta_1,\theta_2\ket_{(\alpha_1,j_1),(\alpha_2,j_2)}^{in} =
|\theta_2,\theta_1\ket_{(\alpha_2,j_2),(\alpha_1,j_1)}^{out}$ for
$j_1\neq j_2$.

The transformation by which the copy numbers are cyclicly permuted
is a symmetry of the multi-copy model. The branch-point
twist field ${\cal T}_n$ is a twist
associated to this symmetry. It is mainly defined by the
equal-time exchange relations \beq
    \Psi_{\alpha,j}(x') {\cal T}_n(x)  = \lt\{ \ba{ll} {\cal T}_n(x) \Psi_{\alpha,j+1}(x')  & x'>x \\ {\cal T}_n(x) \Psi_{\alpha,j}(x') & x'<x \ea \rt.
\eeq (with $\Psi_{\alpha,j+n} \equiv \Psi_{\alpha,j}$).  Other requirements
uniquely fix this twist field: it
is invariant under all other symmetries and has
minimal scaling dimension. As $n\to1$, it becomes the
identity operator.

By symmetry, it is local: $[{\cal T}_n(x), \ep(x')] = 0
\quad (x\neq x')$. Its insertion in a
euclidean correlator produces, as function of the
positions of other fields inserted, a branching by
which the copies are cyclicly connected through the cut extending on its right \footnote{
The shape of the cut does not affect the results.}. Thanks to this branching, the two-point
function is proportional to the partition function on a Riemann
surface, and with an appropriate analytic continuation in $n$, we have
\beq\label{tp}
    S_A = \frc{c}3 \log(\xi_1) + U - \lim_{n\to1} \frc{d}{dn} \bra\vac| {\cal T}_n(r) {\cal T}_n^\dag(0) |\vac\ket~.
\eeq
Below we use locality of branch-point twist fields and the $S$-matrix of the $n$-copy model to justify some of their properties.

{\bf Form factors of branch-point twist fields.}

Let us consider the simple matrix element
\beq
    F_{\mu_1,\mu_2}(\theta_1-\theta_2,n) = \bra\vac|{\cal T}_n(0)|\theta_1,\theta_2\ket_{\mu_1,\mu_2}^{in} \quad (\theta_1>\theta_2)
\eeq where $\mu_i$ are double indices $(\alpha_i,j_i)$. We used
spinlessness and relativistic invariance to write it
as a function of the rapidity difference
$\theta=\theta_1-\theta_2$. This gives the function
for real $\theta>0$ only. Let us now consider its analytic
continuation to complex $\theta$.

For usual local fields, the analytic structure of such matrix elements
is well known \cite{book}. In terms of Mandelstam's $s$ variable,
$s=m_{\alpha_1}^2 +
m_{\alpha_2}^2+2m_{\alpha_1}m_{\alpha_2}\cosh(\theta)$, it has a
branch cut on $s\in[(m_{\alpha_1}+m_{\alpha_2})^2,\infty)$, poles
for $s\in(0,(m_{\alpha_1}+m_{\alpha_2})^2)$ due to bound states,
and no other singularities on the physical sheet, which covers the complex plane excluding the cut $s\in[(m_{\alpha_1}+m_{\alpha_2})^2,\infty)$. Out of integrability, there are other branch points on the cut itself due to inelastic scattering, but these do not affect the physical sheet. On this sheet,
the value on the upper shore of the cut is the form factor with an
$in$ state as above, and that on the lower shore is a form factor
with the same particles and momenta, but forming an $out$ state.

It is a simple exercise to translate this analytic structure in terms of the variable $\theta$. The strip ${\rm Im}(\theta) \in (0,2\pi)$, the physical strip, is a double covering of the
physical sheet (with both shores of the cut separated). The form
factor with $in$ state is on ${\rm Im}(\theta)=0,\,{\rm
Re}(\theta)>0$ and ${\rm Im}(\theta)=2\pi,\,{\rm Re}(\theta)<0$,
and that with $out$ state is on ${\rm Im}(\theta)=0,\,{\rm
Re}(\theta)<0$ and ${\rm Im}(\theta)=2\pi,\,{\rm Re}(\theta)>0$.
Poles are on ${\rm Im}(\theta) \in (0,2\pi),\,{\rm Re}(\theta)=0$,
and are symmetrically distributed about the line ${\rm Im}(\theta)
= i\pi$. Out of integrability, $\theta=0$ and $\theta=2\pi i$ are
branch points, since the proof that they are ordinary point at integrability relies on the lack of particle production.
Also there are more branch points on the lines ${\rm Im}(\theta)=0$ and ${\rm Im}(\theta)=2\pi$ (i.e. outside of the physical strip) at inelastic scattering thresholds, and in general the analytic structure outside of the physical strip on the $\theta$-plane is very complicated.

An important aspect of the present paper compared to previous works at integrability \cite{a,b} is to show that we only need the analytic structure on the physical strip: in particular the inelastic-scattering branch points on its boundary do not affect the derivation.

For the branch-point twist fields ${\cal T}_n$, these properties
are modified. In order to understand them, we will use an
intuitive picture in the euclidean theory, where an imaginary shift of rapidity corresponds to a
rotation of the wave packet (see fig. 1). A complete explanation would require the use of Feynman diagrams, but developing these details is out of the scope of the present paper. However the arguments we present show that the well-known analytic structure described above for ordinary fields is a direct consequence of this intuitive picture. Also, more general analytic properties of twist fields form factors in integrable models were derived following similar ideas in combination with integrability, and could be verified to a large extent thanks to exact solutions. This gives strong support to arguments from this intuitive picture in general. Although twist field form factors were only studied at integrability until now, the arguments, as is clear below, hold also out of integrability for two-particle form factors.
\begin{figure}
\mbox{\includegraphics[width=3cm,height=2cm,angle=0]{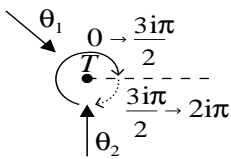}}\vspace{-0.4cm}
\caption{Rotating the $\theta_1$ wave packet in the euclidean
plane. A shift by more than $3i \pi /2$ brings it from copy $j_1$
to copy $j_1+1$ (we chose the branch cut to go straight to the right).}\vspace{-0.5cm}
\end{figure}

First, the value on $\theta<0$ is as usual:
\[
    F_{\mu_1,\mu_2}(\theta_1-\theta_2,n) = \bra\vac|{\cal T}_n(0)|\theta_1,\theta_2\ket_{\mu_1,\mu_2}^{out} \quad (\theta_2>\theta_1)~.
\]
Intuitively, from particles in an $in$-state, continuously
changing $\theta_1-\theta_2$ to make it negative gives particles
travelling away from each other, which is an $out$ state. In the case
where  $j_1\neq j_2$, particles do not interact, so that
\beq\label{free}
    F_{\mu_1,\mu_2}(\theta,n) = F_{\mu_2,\mu_1}(-\theta,n) \quad (j_1\neq j_2)~.
\eeq

Second, the value at $\theta+2\pi i$ for real $\theta$ can be
obtained by rotating the wave packet of particle 1
clockwise. As the rapidity $\theta_1$ arrives at $2\pi i$,
particle 1 is on the right of particle 2 and they are travelling
in the $out$ state configuration. However, particle 1 is now on
copy $j_1+1$: it is the state
$|\theta_2,\theta_1\ket_{\mu_2,\h\mu_1}^{out}$ with $\h\mu_1 =
(\alpha_1,j_1+1\; {\rm mod} \; n)$. Hence we have \beq\label{per}
    F_{\mu_1,\mu_2}(\theta+2\pi i,n) = F_{\mu_2,\h\mu_1}(-\theta,n)~.
\eeq

Third, the function $F_{\mu_1,\mu_2}(\theta,n)$ has poles at
purely imaginary values of $\theta$ for which bound
states of particles $\alpha_1$ and $\alpha_2$ provide additional
on-shell channels. These bound states can
occur only if the particles travel on the
same copy when they interact. Shifting ${\rm Im}(\theta_1)$ by an amount
smaller than $\pi$, particles 1 and 2 interact on
the left of the field ${\cal T}_n$, where they are respectively on
copies $j_1$ and $j_2$. Hence for ${\rm Im}(\theta) \in (0,\pi)$,
poles may occur only for $j_1=j_2$. On the other hand, going
beyond $\pi$, they interact on the right, where they are on copies
$j_1$ and $j_2-1$, or $j_1+1$ and $j_2$. Hence for ${\rm
Im}(\theta) \in (\pi,2\pi)$, poles occur only for $j_1=j_2-1 \ {\rm mod}\; n$.

Finally, there is an additional pole at $\theta=i\pi$, a
``kinematic pole''. Consider $\theta_1\to \theta_1+i\pi$. The
resulting wave packet represents a single-particle $out$ state
with (real) rapidity $\theta_1$ and particle type
$(\b\alpha_1,j_1)$, that is ${\ }^{out}_{\b\alpha_1,j_1}\bra
\theta_1|$, where $\b\alpha_1$ is the anti-particle of $\alpha_1$.
In regions of space where the $in$ and $out$ wave packets
correspond to the same particle type and the same copy, they
overlap. There are two distinct regions where overlaps may occur:
$x\to\infty$ and $x\to-\infty$. They provide the main contribution
to the associated matrix element for $\theta_1\sim\theta_2$:
\[\ba{l}
     \delta_{\b\alpha_1,\alpha_2} \bra{\cal T}_n\ket m \sqrt{\cosh\theta_1\cosh\theta_2}\int dx\, e^{-i xm (\sinh\theta_1-\sinh\theta_2)} \times \\
    \quad\times \lt(\delta_{j_1,j_2} \Theta(-x) + \delta_{j_1+1,j_2} \Theta(x)\rt)
\ea\]
where $\Theta(x)$ is the step function, and Kronecker delta functions of
copy numbers are modulo $n$. This gives both a
Dirac delta function at $\theta_1=\theta_2$, and a pole (its principal
value) as $\theta_1\to\theta_2$. Only the pole can be seen from
the analytic continuation in $\theta_1$:
\beq\label{pol}
    -i F_{\mu_1,\mu_2}(\theta+i\pi) \sim \frc{\delta_{\b\alpha_1,\alpha_2}\bra{\cal T}_n\ket}{\theta}\lt(\delta_{j_1,j_2} - \delta_{j_1+1,j_2}\rt) ~.
\eeq

Apart from all these poles, the form factors are analytic
in the physical strip ${\rm Im}(\theta) \in (0,2\pi)$. Using (\ref{free}) and (\ref{per}),
\beq
    F_{(\alpha_1,j_1),(\alpha_2,j_2)}(\theta) = F_{(\alpha_1,1),(\alpha_2,1)}(\theta+2\pi i (j_1-j_2))
\eeq
for $0\leq j_1 - j_2 \leq n-1$. Then, from the pole conditions and (\ref{pol}),
$F_{(\alpha_1,1),(\alpha_2,1)}(\theta)$ is
analytic in the {\em extended physical strip} ${\rm Im}(\theta)
\in (0,2\pi n)$ except for possible bound-state poles at purely imaginary
$\theta$ on ${\rm Im}(\theta) \in (0,\pi)$ and ${\rm Im}(\theta)
\in (2\pi n-\pi,2\pi n)$, and for kinematic poles, when $\alpha_1=\b\alpha_2$, at $\theta=i\pi$
and $\theta=(2n-1)i\pi$ with residues $i\bra
{\cal T}_n\ket$ and $-i\bra {\cal T}_n\ket$. This is the
structure found in \cite{a,b} in integrable QFT. The larger analytic region of $F_{(\alpha_1,1),(\alpha_2,1)}(\theta)$
is surprising, since for usual local fields in non-integrable QFT there are branch points on ${\rm Im}(\theta)=2\pi$. It was instrumental
that particles on different copies do not interact, and that a shift by $2\pi i$ changes the copy number. Beyond the extended physical strip, the analytic structure is much more complicated.

{\bf Entanglement entropy.}

The two-point function in (\ref{tp}) can be expanded
at large distances by inserting intermediate states between the
twist fields and using space translation covariance. In
particular, the two-particle contribution is:
\beqa
    && {\langle\mathcal{T}_n\rangle^{2}} \frc{n}{8\pi ^2} \sum_{\alpha,\beta=1}^\ell
    \int\limits_{-\infty}^{\infty } \int\limits_{-\infty }^{\infty }d\theta_{1}d\theta _{2} f_{\alpha,\beta}(\theta_{1}-\theta_2,n)\,\times \n
	&& \times\,
    e^{-r(m_\alpha\cosh\theta_1 + m_\beta\cosh \theta_2)}~, \label{pss} \\
	&& \label{fa}
	{\langle\mathcal{T}_n\rangle^{2}}f_{\alpha,\beta}(\theta,n) =
	\sum_{j=0}^{n-1}\left|F_{(\alpha,1)(\beta,1)}(\theta+2\pi i j,n)\right|^2.
\eeqa
In order to evaluate the entanglement entropy, we
need to analytically continue from $n=1,2,3,\ldots$ to $n\in
[1,\infty)$, then to take the derivative at $n=1$. This was done in \cite{a,b} in various ways.
We recall here the main steps, emphasizing that we only need the analytic structure on the extended physical sheet, independently from the structure on its boundary (where there are inelastic-threshold branch points) or further away.

First for $n\in[1,\infty)$,  $F_{(\alpha,1)(\beta,1)}(\theta)$
has poles for $\alpha=\b\beta$ at
$\theta=i\pi$ and $\theta=2i\pi n -i\pi$ with unchanged residues,
and no poles in the strip ${\rm Im}(\theta) \in (\pi,2\pi n-\pi)$.
As noted in \cite{b}, this follows from the picture of
a space with two conical singularities of angle $2\pi n$, valid for real positive n:
kinematic poles for particles going past the conical singularity
on its left and right, and no bound states in the extra space.
Second, form factors vanish as $n\to 1$ since ${\cal T}_n(x)\to{\bf 1}$.
A vanishing like $n-1$ was observed in \cite{a,b}, and
we assume this still holds. This implies no
contribution from one-particle form factors. The two-particle contribution comes from the non-uniform convergence of form factors as $n\to1$,
due to the collision of kinematic poles, as explained in \cite{a}. More precisley, the sum
in (\ref{fa}) can be done by $\sum_{j=0}^{n-1} f(j)
=f(0) + (2i)^{-1} \oint dz f(z) \cot\pi z - Q = f(0) + \int_{-\infty}^\infty dy (f(n+iy-\gamma) \cot\pi (iy-\gamma) - f(iy+\gamma)\cot\pi(iy+\gamma))/2 - Q$
with $0<\gamma<1/2$ and where $Q$ cancels the residues of $f(z)$. Note that the boundary of
the extended physical sheet is avoided.
Then, as $n\to1$ the kinematic residues in $Q$ collide and give \cite{a} $\lt(\frc{\p}{\p n} f_{\alpha,\beta}(\theta,n) \rt)_{n=1} = \frc{\pi^2}2 \delta(\theta) \delta_{\b\alpha,\beta}$ (a formula like this was obtained earlier in the free Dirac model in \cite{cfh}). This gives (\ref{main}).

{\bf Discussion: entangled pairs.}

The most striking feature of (\ref{main}) is its independence from the scattering matrix.
If the entanglement entropy $S_A$ were counting a ``number of links'' connecting points of
$A$ and $\b{A}$, we would have (a regularised version of) $S_A = \int_{A} dx \int_{\b{A}} dx' s(x-x')$
for any region $A$. This agrees with $S_A = S_{\b{A}}$ and holds explicitly in some valence-bond descriptions \cite{vb1,vb2,vb3} (but see \cite{js}).
It is equivalent to saying that the mutual information of \cite{CH04} is extensive, or to $S_A = -\sum_{x<x'\in\partial A} \eta(x)\eta(x') S_{[x,x']}$
where $\eta(x)$ is $\pm1$ if $x$ is a left/right boundary of a
connected component. However, there is convincing numerical evidence that the latter formula fails (although slightly)
in general two-dimensional QFT \cite{chnum}. Yet, factorisation of twist-field correlation
functions immeditaly implies extensivity of the mutual information in the large-distance limit. Hence
the failing should only be due to some ``non-locality'' at the end-points of the ``links'', and we may interpret
$s(x) = -\frc12 \frc{d^2 S_{[0,x]}}{dx^2}$, for $x$ much larger than the correlation length, as measuring an ``entanglement density'', a correlation between quantum disturances a distance $x$ apart.

A natural candidate is from virtual pairs created far in the past (see fig. 2), in a similar way to what happens in quenching \cite{ccq}. Technically, the collision of kinematic poles indeed says that we are considering particles coming from a common virtual pair.
The probability that an entangled pair survives for a time $t$ is essentially ruled
by quantum uncertainty principles, $\propto e^{-tE}$ where $E$ is the total energy, independently from the scattering
matrix. The trajectories are linear on the world sheet since in one dimension, conservation of
momentum and energy forbids smooth changes of their rapidities. Hence the main contribution to $s(x)$ at large distances
should be independent from the scattering matrix. 
\begin{figure}
\mbox{\includegraphics[width=3cm,height=2cm,angle=0]{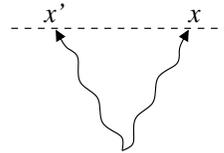}}\vspace{-0.4cm}
\caption{A pair contributing to the entanglement density.}\vspace{-0.5cm}
\end{figure}
For instance, a pair of particles of mass $m$ and velocities $v$ and $-v$ has life-time $E^{-1} = (2m)^{-1}
\sqrt{1-v^2}$. When hitting points $x$ and $x'$, it survived for a time $t=(x-x')/(2v)$.
Its contribution is $m^2 dv \,f(v) e^{-tE}$, which should be positive ($f(v)>0$), and
$s(x-x') \sim  \sum_{\alpha=1}^\ell m_\alpha^2 \int_0^1 dv\, f(v) e^{-\frc{m_\alpha(x-x')}{v\sqrt{1-v^2}}}$.
The exponent is maximum at
$v=1/\sqrt{2}$, where it has value $-2m_\alpha(x-x')$, in agreement
with the correction terms (\ref{main}). Also, (\ref{main}) implies
$f(v)=\frc{1}{32v^3(1-v^2)^2}$, which indeed is positive for
$v\in(0,1)$.

In conclusion, we evaluated the leading large-distance correction terms
for the entanglement entropy in any unitary two-dimensional QFT.
This is one of the few examples of an exact low-energy result, and of the use of the analytic
structure of form factors, out of integrability. The heuristic arguments
we provided give a physical explanation of the main features: independence from the scattering matrix, and why the collision of kinematic poles lead to this result.
It would be useful to have a derivation of the form factor properties from perturbation theory, a better study
of the analytic continuation in $n$ involved, and a better understanding of
extensivity properties of the mutual information.

{\bf Acknowledgments}

I am grateful to J. L. Cardy and O. A. Castro Alvaredo for
discussions and comments about the manuscript, and to H. Casini and
M. Huerta for sharing their insights.

\null

\end{document}